\documentstyle[prl,aps,epsf,twocolumn]{revtex} 

\newcommand{\be}{\begin{equation}} 
\newcommand{\ee}{\end{equation}} 
\newcommand{\bdis}{\begin{displaymath}} 
\newcommand{\edis}{\end{displaymath}} 
 
\newcommand{\ie}{{\it i.e.}} 
\newcommand{\etal}{{\it et al.}} 
\newcommand{\la}{\left\langle} 
\newcommand{\ra}{\right\rangle}
\newcommand{\lp}{\left(} 
\newcommand{\rp}{\right)}
\newcommand{\rms}{r.m.s.}

\begin{document} 
%
%
\bibliographystyle{prsty}

\title{Intermittency and structure functions in channel flow turbulence} 

\author{F. Toschi$^{1,2}$, G. Amati$^3$, S. Succi$^{4}$, R. Benzi$^{5}$
and R. Piva$^{6}$ } 
\address{$^{1}$  Dipartimento di Fisica, Universit\'a di Pisa, P.zza
Torricelli 2, I-56100 , Pisa, Italy.\\ 
$^{2}$  INFM, Unit\`{a} di Tor Vergata, Roma.\\ 
$^{3}$ CASPUR, Universit\`a ``La Sapienza'', P.le Aldo Moro 5, I-00185,
Roma, Italy.\\ 
$^{4}$ Istituto Applicazioni Calcolo ``Mauro Picone'', V.le Policlinico
137, I-00161, Roma , Italy.\\ 
$^{5}$ AIPA, Via Po 14, I-00100, Roma, Italy.\\ 
$^{6}$ Dip. di Meccanica e Aereonautica, Universit\`a ``La Sapienza'',
Via Eudossiana 18, I-00184, Roma, Italy. } 
\date{\today} 
\maketitle 

\begin{abstract} 
We present a study of intermittency in a turbulent channel flow. 
Scaling exponents of longitudinal streamwise structure functions, 
$\zeta_p /\zeta_3 $, are used as quantitative indicators of intermittency.

We find that, near the center of the channel the values of 
$\zeta_p /\zeta_3 $ up to $p=7$ are consistent with the assumption of 
homogeneous/isotropic turbulence. 
Moving towards the boundaries, we observe a growth of intermittency which 
appears to be related to an intensified presence of ordered vortical 
structures. 
In fact, the behaviour along the normal-to-wall direction of suitably 
normalized scaling exponents shows a remarkable correlation with the 
local strength of the Reynolds stress and with the \rms\ value of helicity 
density fluctuations. 
We argue that the clear transition in the nature of intermittency 
appearing in the region close to the wall, is related to a new 
length scale which becomes the relevant one for scaling in high shear
flows. 
\end{abstract}
 
\narrowtext

Spatio temporal intermittency is one of the most intriguing properties of 
fluid dynamics turbulence. 
Intermittency can be described by means of the scaling behavior 
of the structure functions $S_p(r)$ built out of the velocity difference, 
namely $S_p(r)=\la\delta v(r)^p \ra$ where $ \delta v(r) = \delta
\vec{v}(\vec{r}) 
\cdot \vec{r}/r $. 
In the inertial range, $S_p(r)$ scale homogeneously with exponents 
$\zeta_p$, \ie\ $S_p(r) \propto r^{\zeta_p}$. 
Intermittency reflects in anomalous values 
(\ie\ different from the Kolmogorov 
1941 prediction $\zeta_p = p/3$) of the 
$\zeta_p$ (see Frisch \cite{frish}). 
Although many efforts have been devoted to the 
understanding of intermittency in 
homogeneous and isotropic turbulence, the case of 
wall turbulence dominated by very strong shear flow 
is still under debate. 
In fact, while the decrease of scaling exponents 
towards the wall has been recently pointed out 
by both experimental and numerical investigations 
\cite{etc7,torino}, a physical explanation 
of this effect is still missing.\\

The main objective of this paper is to analyse 
the scaling exponents (if any) and to explore their 
functional dependence on the non-homogeneous coordinate. 
To this purpose, we investigate numerically a 
channel flow \cite{KMM,JiMo}, possibly the 
simplest instance of a shear-dominated flow.  

The problem of characterizing the complex 
phenomena arising in the near-wall region, and their 
relation to the lack of isotropy 
has been analyzed in depth by several authors 
({\it e.g.}, see Antonia \etal \cite{Antonia1} and L'vov and Procaccia \cite{PROCA}). 
With specific reference to intermittency, Kutznetsov \etal 
\cite{kuz} presented an experimental investigation of fine-scale 
structure of turbulence for different shear-flows. 
Even though they recognize that scaling exponents 
may be different for various flows or various locations 
of the same flow, they do not seem to address the issue 
of the spatial dependence of the exponents $\zeta_p$. 
This issue is examined in a recent paper 
by Camussi \etal \cite{camussi}, 
in which the authors propose a technique to identify 
coherent structures and relate them to the increase 
of intermittency in (homogeneus and non-homogeneus) 
grid turbulence. 
The present work is fairly distinct in purpose, since 
we analyse the spatial behaviour of the scaling exponents 
and their link to coarse-grained features of the 
flow in the near-wall region. 
Besides the fundamental interest on its own, the existence of such 
correlation could prove very valuable for the design of more efficent 
Large-Eddy-Simulation models (see for instance Scotti and Menevau 
\cite{MENEVAU}).\\ 

We have performed a direct numerical simulation achieving a 
high statistical accuracy 
(about $10^3$ in time units $U_0/h$, where $U_0$ is 
the centerline velocity and $h$ is the channel half-width). 
Numerical simulations have been performed on a massively parallel 
machine using a LBE (Lattice Boltzmann Equation) code. 
The spatial resolution of the simulation was $256\times 128\times 128$ grid points. 
Periodic boundary conditions were imposed along the streamwise ($x$) and 
spanwise ($z$) directions, whereas no-slip boundary conditions 
were applied at the top and the bottom planes (normal to wall direction, $y$). 
The Reynolds number is $Re \simeq 3000$. 
Further details about the numerical scheme can be found in \cite{giorgio} 
and references therein. 
In the following we use wall units defined as $y^+=y \cdot v^*/\nu$ and $v^+=
v/v^*$ where $v^*$ is the friction velocity \cite{landau}. 
In these units, the channel is $640 $ long, $320$ wide and $ 320$ high.\\

To study intermittency in the channel, we introduce the following 
$y$-dependent longitudinal streamwise structure functions: 
\nobreak
\begin{eqnarray}
\nonumber &&S_p(r^+,y^+)=\la\left| (v_x(x^+ + r^+,y^+,z^+) -v_x(x^+,y^+,z^+)\right|^p\ra\\
&&
\end{eqnarray} 
\narrowtext
The average is taken at a fixed $y^+$ value 
(the normal to wall coordinate). 
The quantities $S_p(r^+,y^+)$ have 
been measured for each value of $y$. 
Our data set allows enough statistical accuracy to estimate 
$S_p (r^+,y^+)$ for $p \le 7$. 
Due to the low Reynolds number, we use ESS \cite{ess} in order to extract 
$\zeta_p$ values. 
We remind that ESS consists of measuring structure functions as a function 
of $S_3$ rather than in terms of space separation $r$. 
This procedure allows a much better accuracy for the evaluation 
of the scaling exponents, although it does not provide any estimate of the
$y^+$-dependence 
of $\zeta_3$. 
In order to compute the scaling exponents $\zeta_p(y)$ we have analyzed
the 
ESS local slopes $ D_{p,q}(r^+,y^+) = d\log(S_p(r^+,y^+)) /
d\log(S_q(r^+,y^+))$ 
for each value of the $ y^+ $ coordinate. 
We have found two regions in $y^+$, hereafter referred to as region 
H ("Homogeneous") and region B ("Boundary") respectively, where well 
defined constant local slopes for 
the scaling exponents can be detected. 
Region H is close to the center of the channel ($y^+ \ge 100$) while
region 
B is close to the viscous sublayer ($20 \le y^+ \le 50$). 
In region H, the scaling exponents $\zeta_p(H)$ are found to be 
approximately the same as the ones measured in homogeneous and 
isotropic turbulence. 
On the other hand, in region B the scaling exponents $\zeta_p(B)$ have 
been found to be much smaller than $\zeta_p(H)$. 
Moreover, while in region H the scaling range  starts at 
$r^+ \ge 25$, in region B the scaling range starts 
at $r^+ \ge 50$, consistently with previous findings \cite{PRE}. 
In the intermediate region between region H and region B, 
it is difficult to identify a range in $r$ where a scaling 
exponent can be defined with enough confidence.\\ 
%

In order to clarify the discussion, we show in Fig. \ref{fig1a} and 
Fig. \ref{fig1b} the local slopes 
$D_{6,3}(r^+,y^+) $ and $ D_{4,2}(r^+,y^+)$ 
respectively for $y^+ = 30,70,80,150$. 
In the intermediate region (\ie\ $y^+ =70,80$), the analysis 
in terms of local slope does not provide a well defined scaling exponent 
since the plateau in $ r^+ $ is very short. 
In Fig. \ref{fig2a} and Fig. \ref{fig2b} we show $\zeta_6 /\zeta_3$ and 
$\zeta_4 /\zeta_2$ respectively as a function of $ y^+ $  with the 
associated error bars. 
The large error bars in the region $50 \le y^+  \le 100$ indicate that 
scaling exponents defined through $D_{p,q}(r,y^+) $ 
are poorly defined and should be considered just as effective exponents 
obtained by the power law fit of the ESS analysis. 
The situation described in Fig. \ref{fig1a} and 
Fig. \ref{fig2a} is similar for 
all the scaling exponents $\zeta_p$ computed in our analysis. 
Finally, in Table \ref{table1}, we list the numerical values of 
the scaling exponents for region B, region 
H and for the homogenous isotropic turbulence \cite{ESS}. 
Our results indicate that there is a 
transition in the nature 
of intermittency between region H and region B. 
While in region H intermittency is close to what has been observed 
in isotropic and homogenous turbulence, in region 
B much stronger intermittency is observed,  which 
reflects in lower values of the scaling exponents 
$\zeta_p$ for $p >3$ and larger ones for $p<3$. 
Inbetween the two regions, a competition between the 
two types of intermittency 
should take place, leading to a poorly defined scaling law.\\ 
An important question to be addressed concerns the physical 
mechanisms which produce much stronger intermittency 
in region B with respect to region H. 
A preliminary answer to this question is given in Fig. \ref{fig3} where 
the momentum flux $\la v_x' v_y' \ra $ (a particular component of the Reynolds stress tensor) normalized by its 
maximum $\la v_x' v_y' \ra_M $, is shown as a function of $y^+$. 
The momentum flux has a peak within the region B while 
it goes to zero in region H. 
This behaviour clearly indicates that 
intermittency grows dramatically 
in region characterized by strong mean 
shear (strong momentum fluxes). 
The link between momentum flux and the increase of 
intermittency can be  further investigated 
by looking at the inset of 
Fig. \ref{fig3}  where  we represent $\zeta_6(y)/\zeta_3(y)$ from Fig. 
\ref{fig2a} and the rescaled expression 

\begin{equation} 
{{\zeta_6 (y^+)} \over {\zeta_3(y^+)}} - {{\zeta_6(H)} \over {\zeta_3(H)}} = \lp{{\zeta_6(B)} \over {\zeta_3(B)}}-{{\zeta_6(H)} \over {\zeta_3(H)}}\rp 
\frac{\la v_x' v_y'\ra}{\la v_x' v_y' \ra_M}. 
\end{equation} 

From Fig. \ref{fig2a}, we can argue that the increase of intermittency should
be related to the increase of momentum flux and, therefore, to the mean (local)
shear.
 \\ 
 Moreover, it is well known that turbulent flows 
near the wall are characterized by well defined coherent structures. 
In Fig. \ref{fig3} we show the \rms\ helicity $h$ 
($ h_{rms} =\la(\vec{\omega} \cdot \vec{v})\ra_{rms}$) as a function of $
y^+$ 
which is again peaked in region B. 
Coherent structures carry a significant amount of 
helicity while dissipation is found to peak in the interstitial region 
between helicity-carrying structures \cite{amaticasciola}. 
Indeed a clearcut anticorrelation between helicity fluctuations 
and dissipation is systematically detected in our numerical simulations. 
Thus, the alternate presence of regions of high helicity and regions of 
high dissipation may be responsible for the enhancement of intermittency
in 
region B.\\ 
A more quantitative way to investigate the increase of intermittency in 
region B,  can be achieved by the following argument. 
According to Howarth-Von Karman-Kolmogorov equation for homogeneous shear 
flows turbulence (see Hinze \cite{HINZE}), one can define 
a length scale $L_s(y)$ in terms of the mean energy dissipation 
$\epsilon(y)$ and the mean shear $\Sigma(y)$, as follows: 
\begin{equation} 
L_s(y) = \lp{{\epsilon(y)} \over {\Sigma(y)^3}}\rp^{1/2} 
\end{equation} 
In presence of mean shear $\Sigma$, for any scale $r$ we can 
define two characteristic timescales, namely the eddy turnover time 
$r/\delta v(r)$ and $1/\Sigma$. 
We expect that when the mean shear is large enough, 
the eddy turnover time is not the relevant timescale 
for energy transfer from large to small scales. 
The inequality $ r/\delta v(r) < 1/\Sigma $ gives the range of 
scales $r$ where the effect of shear should not be relevant to small scale
statistics.   By using the Kolmogorov estimate $\delta v(r) \sim
\epsilon^{1/3} r^{1/3}$, 
we find that the above inequality can be written as $ r \le L_s $. 
Thus, for $r \le L_s $ one expects that the scaling properties of
turbulence 
are not affected by the mean shear. On the other hand, for $ r \ge L_s$
 one expects that the mean shear may significantly change the amount of
intermittency.
In our numerical simulation $L_s(y)$ becomes small only in the region B, 
\ie\ in the region where an increase of intermittency is observed.  
This result confirms our finding of a rather clear transition in the 
physical nature of intermittency, somehow similar to the Bolgiano 
scaling appearing in the thermal turbulence.\\ 
Hence, this preliminary analysis seems to indicate 
that the change of scaling exponents cannot be reduced to a 
perturbative effect in terms of the mean shear. 
As a final observation, by inspecting the peak value approached by the 
$\zeta_p(y)$ exponents near the wall, we find an interesting similarity
with the values, $\zeta_p^{\mbox{\tiny PS}}$, pertaining to a passive scalar advected by a turbulent 3D homogeneous/isotropic velocity field \cite{sergio}. 
Values of passive scalar $\zeta_p^{\mbox{\tiny PS}}$ are shown together with 
our present data in Table \ref{table1}. 

Table \ref{table1} suggests that the passive scalar behaviour can be traced to these 
helicity-carrying coherent structures being passively advected by the
flow. 
This observation is in qualitative agreement with the results reported by 
Pumir and Shraiman \cite{pumir}.

In conclusion, we have rather clear evidence that, in wall bounded
turbulence, the increase of intermittency near the wall is strongly 
related to the increase of the mean shear.
We have introduced a characteristic length scale $L_s$ induced by the mean
shear whose physical
meaning is equivalent to the Bolgiano scale for natural convection. Velocity
fluctuations at scales $r \ge L_s$ are observed to be more intermittent than 
in homogeneous and isotropic turbulence.

\acknowledgments
The authors would like to thank C. Casciola for useful hints and
suggestions. 
F.T. would like to acknowledge S. Ciliberto for interesting discussions
and for his kind hospitality at ENS-Lyon. 
This work was partially supported by INFM.

\begin{table}
\caption{{Normalized scaling exponents in regions H and B for the present simulation, as compared with the values for homogeneous isotropic turbulence $\zeta_p^{\mbox{\tiny hom}}$ and for a passive scalar $\zeta_p^{\mbox{\tiny ps}}$.}\label{table1}}
\begin{tabular}{cccccc} 
$p$  
      & $ (\zeta_p /\zeta_3)^{\mbox{\tiny hom}} $  
      & $ (\zeta_p /\zeta_3)^{\mbox{\tiny H}} $  
      & $ (\zeta_p/\zeta_3)^{\mbox{\tiny B}}  $ 
      & $ \zeta_p^{\mbox{\tiny PS}}  $  
      & $ \zeta_p^{\mbox{\tiny PS}}/\zeta_3^{\mbox{\tiny PS}}$   \\ 
&&&&&\\ \tableline
1     &0.36 &  0.37  & 0.44 &  0.37  &  0.46   \\ \hline 
2     &0.70 &  0.70  & 0.77 &  0.62  &  0.77   \\ \hline 
3     &1.00 &  1.00  & 1.00 &  0.80  &  1.00   \\ \hline 
4     &1.28 &  1.28  & 1.17 &  0.94  &  1.17   \\ \hline 
5     &1.54 &  1.54  & 1.31 &  1.04  &  1.30   \\ \hline 
6     &1.78 &  1.78  & 1.44 &  1.12  &  1.40   \\ \hline 
7     &2.00 &  2.00  & 1.55 &  1.20  &  1.50   \\ 
\end{tabular}
\end{table}

\newpage 
%
%
\begin{figure}[h] 
\begin{center} 
\caption{Local slope $D_{6,3}$ for four different values of $y^+$. The straight lines represent the fit for H and B regions.} 
\label{fig1a} 
\end{center} 
\end{figure} 
\begin{figure}[h] 
\begin{center} 
\caption{Local slope $ D_{4,2}$  for four different values of $y^+$. The straight lines represent the fit for H and B regions.} 
\label{fig1b} 
\end{center} 
\end{figure} 
\begin{figure}[h] 
\begin{center} 
\caption{Scaling exponents $\zeta_6/\zeta_3$ as a function of $y^+$.
The straight lines represent the fit for H and B regions.} 
\label{fig2a} 
\end{center} 
\end{figure} 
\begin{figure}[h] 
\begin{center} 
\caption{Scaling exponents $\zeta_4/\zeta_2$ as a function of $y^+$.
The straight lines represent the fit for H and B regions.} 
\label{fig2b} 
\end{center} 
\end{figure} 
\begin{figure}[h] 
\begin{center} 
\caption{
The momentum flux $\la v_x v_y\ra$ (dashed line), normalized by its maximum,
is presented together with normalized dissipation (straight line) and 
 \rms\ helicity (dotted line). In the inset the scaling exponent $\zeta_6/\zeta_3$ is presented together with the rescaled momentum flux as defined by equation 2 (continuous line).} 
\label{fig3} 
\end{center} 
\end{figure} 
\begin{figure}[h] 
\begin{center} 
\caption{The characteristic scale $L_s(y^+)$ is presented as a function of $y^+$ (continuous line). The dotted line is reported to enlight the growth of $L_s$ out of the wall region.} 
\label{fig4} 
\end{center} 
\end{figure} 
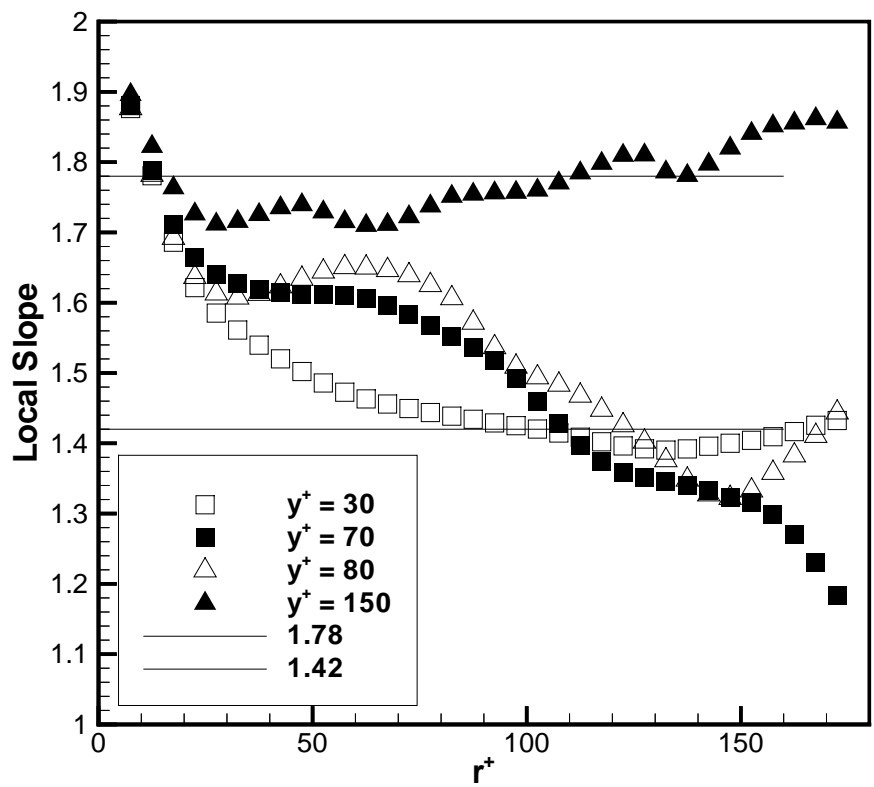
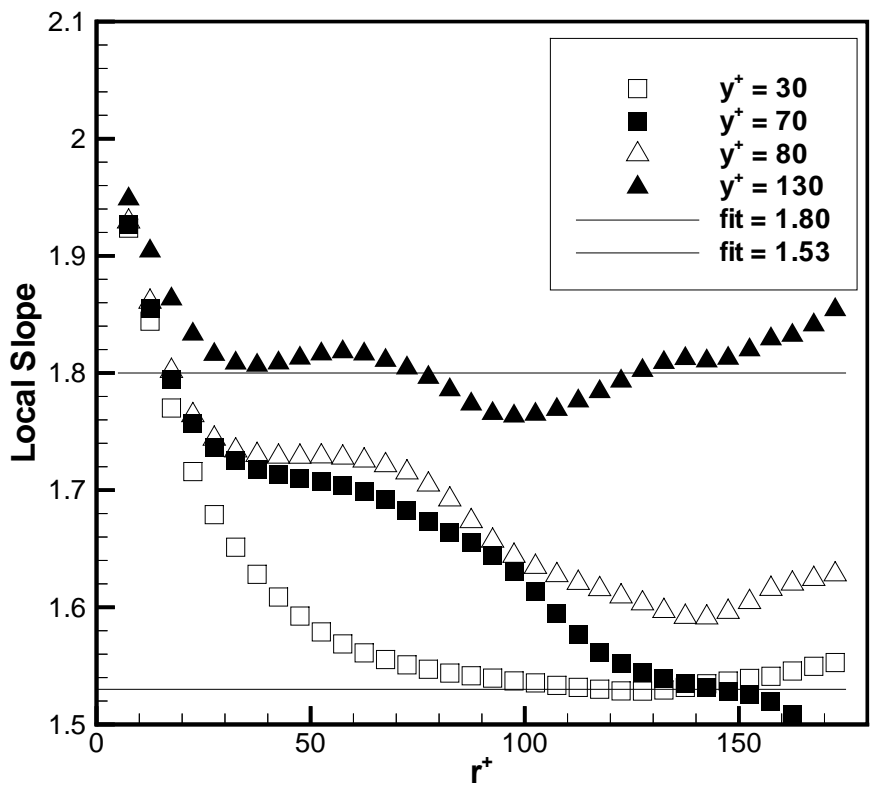
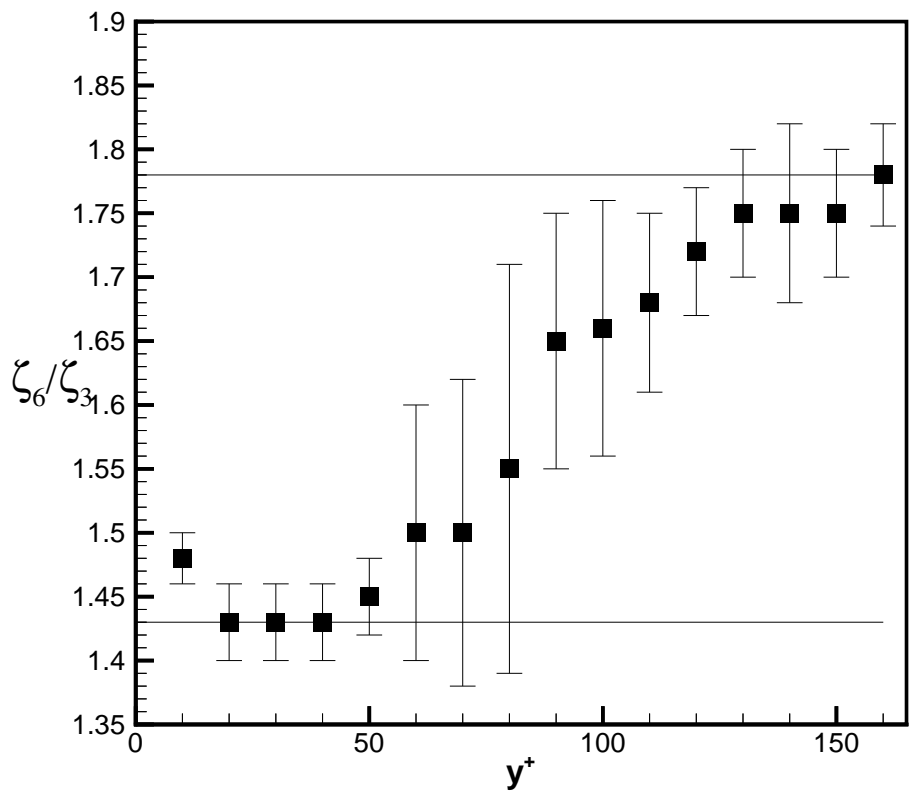
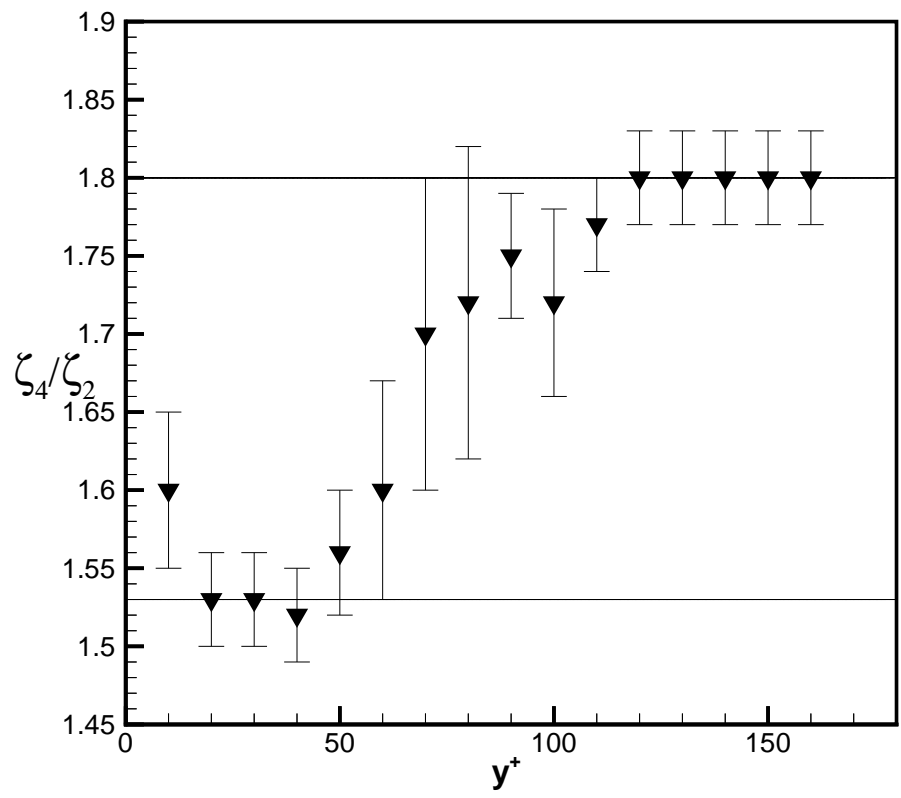
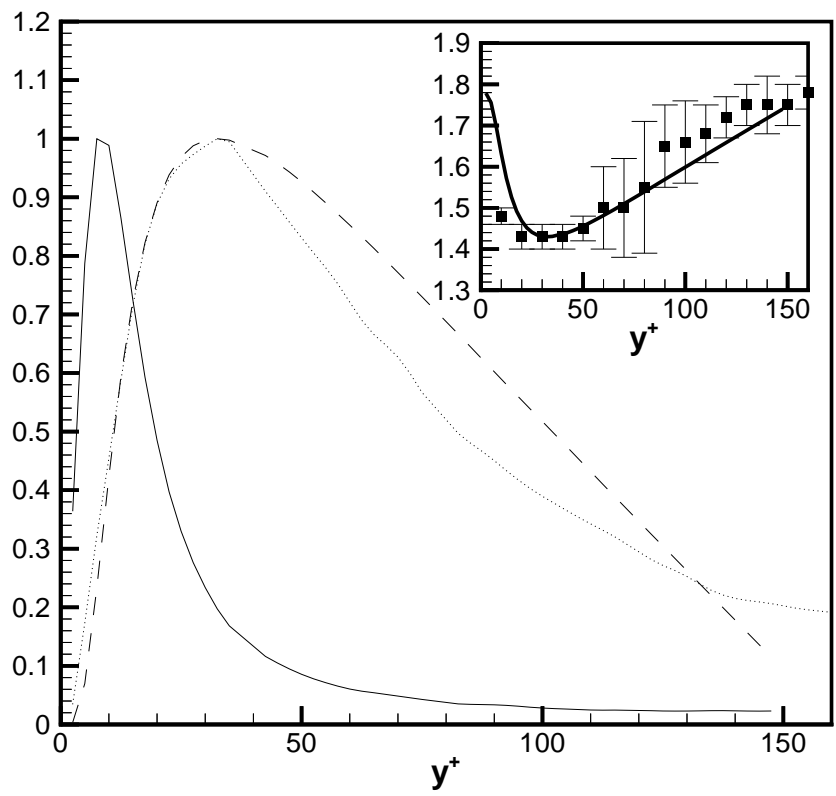
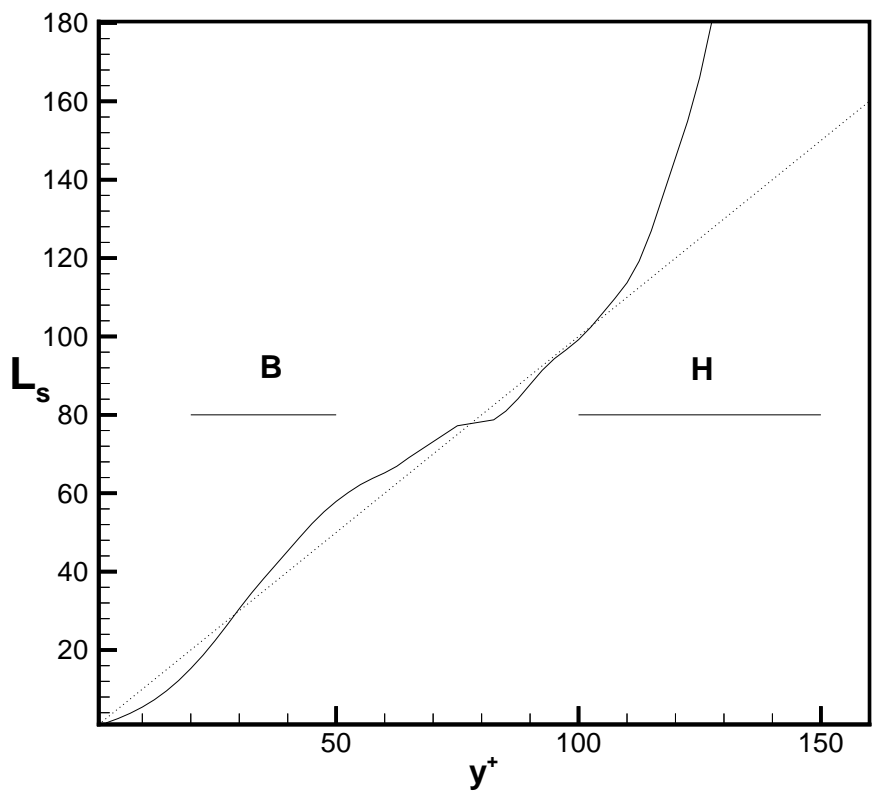
\end{document}